# Integrated Geostationary Solar Energetic Particle Events Catalog: GSEP

Sumanth Rotti ,[1] Berkay Aydin ,[2] Manolis K. Georgoulis ,[3] and Petrus C. Martens[1]

[1]*Georgia State University*
*Department of Physics and Astronomy*
*Atlanta, GA, USA*
[2]*Georgia State University*
*Department of Computer Science*
*Atlanta, GA, USA*
[3]*Academy of Athens*
*Research Center for Astronomy and Applied Mathematics*
*Athens, Greece*

## ABSTRACT

We present a catalog of solar energetic particle (SEP) events covering solar cycles 22, 23 and 24. We correlate and integrate three existing catalogs based on Geostationary Operational Environmental Satellite (GOES) integral proton flux data. We visually verified and labeled each event in the catalog to provide a homogenized data set. We have identified a total of 341 SEP events of which 245 cross the space weather prediction center (SWPC) threshold of a significant proton event. The metadata consists of physical parameters and observables concerning the possible source solar eruptions, namely flares and coronal mass ejections for each event. The sliced time series data of each event, along with intensity profiles of proton fluxes in several energy bands, have been made publicly available. This data set enables researchers in machine learning (ML) and statistical analysis to understand the SEPs and the source eruption characteristics useful for space weather prediction.

*Keywords:* Sun: Solar Energetic Particle (SEP) Events — Solar Proton Event Catalogs

## 1. INTRODUCTION

Solar Energetic Particle (SEP) events are radiation storms of particle fluxes comprising electrons, protons, and heavier ions from the Sun. SEP events are known to originate in large eruptions such as solar flares (SFs) and coronal mass ejections (CMEs) (Reames 1999, 2013; Desai & Giacalone 2016). The number of SEP events occurring in any solar cycle (SC) varies and is much less compared to that of SFs and CMEs, because of the acute directionality effects of SEPs and the fact that they are only detected in-situ (Klein & Trottet 2001; Klein & Dalla 2017; Anastasiadis et al. 2019). The time intensities of particle fluxes are used to define and characterize SEP events. Enhancement above a nominal background level is considered to indicate a possible event. Such time profiles can be used to distinguish the source event as the temporal behavior appears to be different.

The energy of particles in large SEP proton events can reach giga electron volt (GeV) in some instances (Reames 2001; Bruno et al. 2018), and these events can last from a few hours to several days (Kallenrode 2003; Klein & Psoner 2005; Kahler 2005; Cane & Lario 2006). These events have the capacity to disrupt spacecraft operations (Smart & Shea 1992; Pulkkinen 2007), and pose a hazard of radiation exposure to astronauts and aircraft travel in polar routes where protection/shielding is limited (Beck et al. 2005; Schrijver & Siscoe 2010; Schwadron et al. 2010; Jiggens et al. 2019). Understanding the origin and propagation of SEPs is a formidable scientific challenge, and of crucial importance to space weather research (Jackman & McPeters 1987; Gopalswamy 2003). In addition, as these hazards impose significant constraints on space-based activities for both humans and electronic equipment, predicting the event occurrences along with sufficient advanced warning time is of vital importance to operations.

Multiple space and ground-based missions currently obtain in situ solar particle composition and energy spectra fluxes. Researchers have prepared catalogs of SEP events using the available in-situ particle data. Flux measurements beyond the Earth's magnetosphere have been used to populate these catalogs. For example, Wind spacecraft data are used by Kahler (2005) and Miteva et al. (2018). Solar and Heliospheric Observatory (SOHO) mission data are used by Cane et al. (2010) and Paassilta et al. (2017). SEPServer (Vainio et al. 2013) uses data from Wind, SOHO and the Advanced Composition Explorer (ACE).

Solar proton event catalogs based on near-Earth observations such Geostationary Operational Environmental Satellite (GOES) and Interplanetary Monitoring Platform (IMP-8) are of interest in this study. In Table 1, a list of existing SEP event catalogs utilizing near-Earth satellites is presented.

Researchers who do not often work on data processing issues can benefit enormously from a carefully integrated dataset in testing theoretical or working hypotheses. Whether we consider data from active regions or time series measurements or images of the full solar disk, cleaned and organized datasets are crucial when it comes to building space weather forecasting systems. To implement machine learning (ML) methodologies, cleaned datasets are vital during research phase because unavailability of feasible data creates short-fall to research-oriented approaches. Hence, careful integration of observational evidences backed with theoretical reasoning are necessary while developing datasets. In addition, it is necessary to identify and correct errors, shortcomings and caveats in the measurements and corresponding metadata because (1) data quality can impact the research output; and (2) it can mislead both model and data-driven analysis. To bridge the gap, comparisons and integration of data catalogs are critical for improving the performance of event predictions and the outputs of comparative scientific studies. The Astroinformatics cluster at Georgia State University pursues data-driven research with particular solar-physics applications (Angryk et al. 2020; Rotti et al. 2020). One of the areas is SEP event forecasting. The tasks include integrating reference data sets, constructing metadata with well-defined statistical parameters derived from the measurements, and post-processing. Efforts on SEP event predictions using ML have been going on since the last decade (Laurenza et al. 2009; Falconer et al. 2011; Engell et al. 2017; Stumpo et al. 2021; Kasapis et al. 2022). The GSEP dataset to be discussed in this work supports the SEP predictions research area in two key perspectives, namely, in providing:

- Metadata for the source active regions, associated flares, CMEs and radio bursts.

- Time series subsets of proton fluxes for the SEP event duration with an observation window of 12 hours.

This paper aims to bring together available SEP event catalogs based on GOES data as explained in Section 3. We have integrated a comprehensive list of SEP events with reference to their parent SFs, and CMEs. The database comprises of 341 SEP events, extending from 1986 to 2017. Section 4 describes the processes undertaken in data retrieval, pre-processing the GOES data, and generation of the catalog in discussion. In Section 5, the results with observational details and minor differences between the catalogs are summarized. The purpose of this work is to provide the largest possible base for experimenting with statistical and machine learning models on SEPs and their solar source (SF and CME) properties. Source eruptions can then be correlated with photospheric magnetic field and metadata thereof to complement and physically / statistically connect SEP events with their solar, low-atmospheric progenitors.

2. BACKGROUND

The National Oceanic and Atmospheric Administration (NOAA) continuously monitors the near-Earth space environment through GOES in geostationary orbit (Sauer 1989; Bornmann 1996). The GOES satellites record the solar activity and the in-situ radiation environment. They usually operate in pairs with one satellite over the West coast and another over the East coast of the United States in geostationary orbit. NOAA classifies the two GOES satellites making parallel measurements as the "primary" and the "secondary" one. Over the three SCs from 1986 to 2017, eleven different GOES satellites have been launched and commissioned.

The GOES series carries various instruments, including the Space Environment Monitor (SEM, Grubb 1975). One of its constituent detectors is called the Energetic Particle Sensor (EPS, Onsager et al. 1996) on GOES-5 to 12. The twin EPS system on GOES-13 to 15 is called the Energetic Proton, Electron, and Alpha Detector (EPEAD). There are seven proton channels in the EPS/EPEAD taking in-situ differential measurements with characteristic energies spanning from a few up to several hundreds of megaelectron volts (MeVs) (Sandberg et al. 2014). Furthermore, these channels are binned to seven nominal integral energies: P1 ($>$1 MeV), P2 ($>$5 MeV), P3 ($>$10 MeV), P4 ($>$30 MeV), P5 ($>$50 MeV), P6 ($>$60 MeV), and P7 ($>$100 MeV). However, GOES-09 & 14 missions, and channels P6 & P7 on

**Table 1.** The consulted list of SEP catalogs based on the GOES data.

| Catalog | Period | Threshold | | Solar source | | |
| --- | --- | --- | --- | --- | --- | --- |
| | | Channel (MeV) | Intensity (pfu) | Flare | CME | Active Region |
| Kurt et al. (2004) | 1970 - 2002 (253) | >10 | >10 | Y | N | Y |
| Belov et al. (2005) | 1975 - 2003 (1144) | >10 | >0.1 | Y | Y | Y |
| Gerontidou et al. (2009) | 1996 - 2006 (368) | >10 | >0.1 | Y | Y | N |
| Dierckxsens et al. (2015) | 1997 - 2006 (90) | >10 | >0.1 | Y | Y | N |
| Papaioannou et al. (2016) | 1984 - 2013 (314) | >10 | >0.5 | Y | Y | Y |
| PPS (Kahler et al. 2017) | 1986 - 2016 (138) | >50 | >1.0 | Y | N | N |
| CDAW-SEP[a] | 1998 - 2017 (152) | >10 | >10 | Y | Y | Y |
| NOAA-SEP[b] | 1976 onwards (266) | >10 | >10 | Y | Y | Y |
| RF-SPE[c] | 1970 - 2019 | >10 | >1.0 | Y | Y | Y |

NOTE— The value in parenthesis under 'Period' denotes the number of events reported in that catalog.
[a] https://cdaw.gsfc.nasa.gov/CME_list/sepe/
[b] https://umbra.nascom.nasa.gov/SEP/
[c] http://www.wdcb.ru/stp/solar/solar_proton_events.html

GOES-12 have failed (Rodriguez et al. 2014). Nonetheless, measurements are available from as many as nine GOES satellites, from GOES-05 to GOES-15.

NOAA's Space Weather Prediction Center (SWPC) provides radiation storm products based on proton intensity levels as observed by SEM's particle sensors (Rodriguez et al. 2014; Kress et al. 2020). The severity of the proton events is measured using the NOAA Solar Radiation Storm Scale (S-scale). SWPC's S-Scale relates to biological impacts and effects on technological systems. The S-scale relies on the ≥10 MeV integral peak proton flux that characterizes an SEP event's 'size' or intensity, although different peak fluxes logarithmically define different event sizes. The base threshold, associated with a S1 storm, corresponds to a GOES 5-min averaged ≥10 MeV integral proton flux exceeding 10 particle flux units (1 pfu = 1 particle/$cm^2$ sr s) for at least three consecutive readings (Bain et al. 2021). As can be seen in Table 1, many studies do not always conform to this definition because multiple enhancements or rises in the proton flux are considered in one SEP event. Differences in event definition occur due to different needs in research and operations, making it hard to achieve a harmonized data treatment.

## 3. SOURCE CATALOGS

We consider three SEP event catalogs developed using GOES data as sources: PSEP (Papaioannou et al. 2016), CDAW-SEP[1] and NOAA-SEP[2]. We classify the former two as "primary" and the latter as "reference" data.

### 3.1. *PSEP Catalog*

Papaioannou et al. (2016) developed a catalog of 314 well-defined SEP events by statistically studying the relationship between SEP events and possible source eruptions such as flares and CMEs. For each event, they calculated the SEP onset times per event and per channel using the so-called *σ method* (for details, see Papaioannou et al. (2014)). This catalog is based on cleaned differential proton fluxes[3] from EPS made available directly by the the Solar Energetic Particle Environment Modelling (SEPEM) Team (Crosby et al. 2015). The cleaned EPS data set spans over 40 years (1974-2016) and has been cross-calibrated by Sandberg et al. (2014) with data from the Goddard medium energy (GME) instrument on IMP-8. Papaioannou et al. (2016) define an SEP event based on the following threshold parameters:

1. A threshold of 0.01 particles/$cm^2$ sr s MeV (differential flux) above which a possible enhancement was marked.

---

[1] https://cdaw.gsfc.nasa.gov/CME_list/sepe/
[2] ftp://ftp.swpc.noaa.gov/pub/indices/SPE.txt
[3] SEPEM Reference Data Set (RDS): http://sepem.eu/help/SEPEM_RDS_v2-01.zip

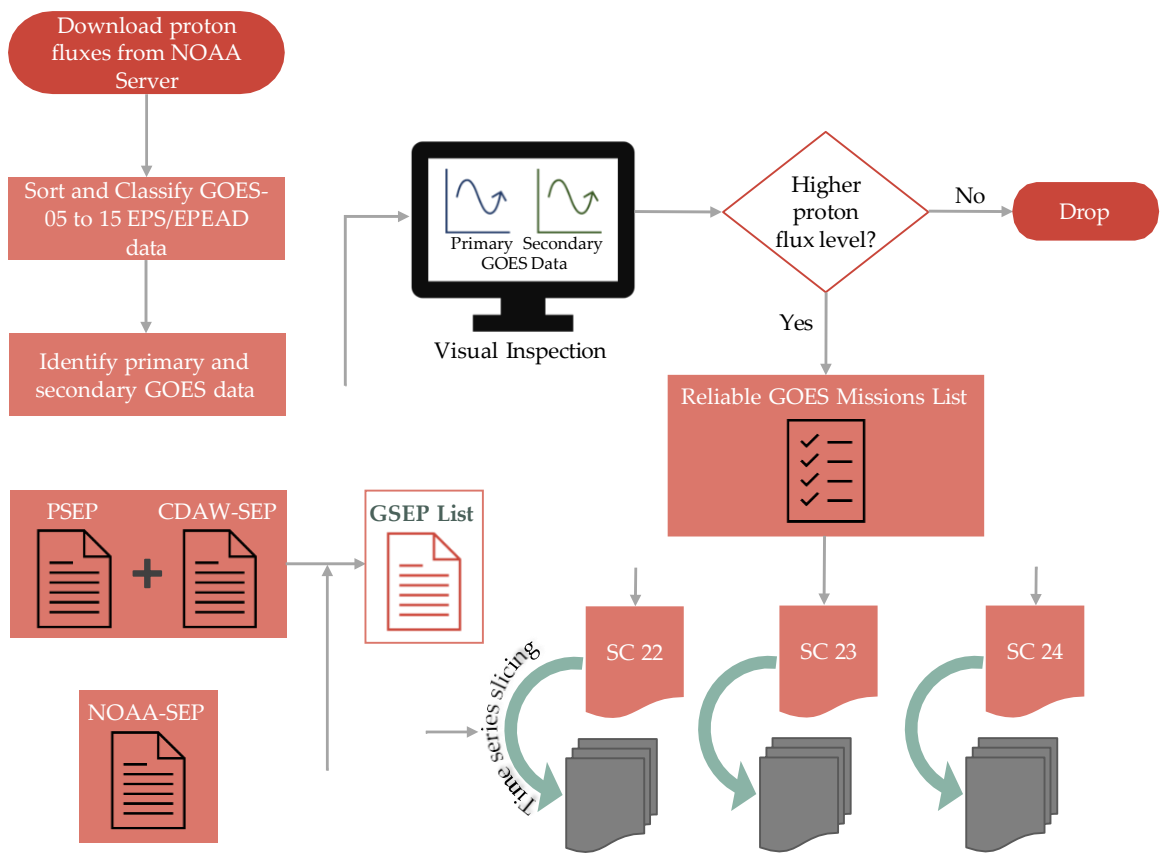

**Figure 1.** Process flow diagram indicating the background work of GOES data inspection, verification, and slicing in this work. The time series slices are generated from the GSEP metadata. For details, see text.

2. A minimum peak of 0.5 pfu of the candidate event.

3. A waiting time of 2 hours between two consecutive candidate events.

4. A minimum event duration of 2 hours.

### 3.2. *CDAW-SEP Catalog*

The series of Coordinated Data Analysis Workshops (CDAW) was organized to analyze the set of all major SEP events (>10 MeV protons crossing the ≥10 pfu threshold) detected by NOAA's GOES spacecraft (Gopalswamy et al. 2002; 2003b). The CDAW-SEP list has 152 events from 1997 to 2017 identified using integral proton data. The only criterion used for event selection was the peak proton flux crossing 10 pfu in the 10 MeV channel following the NOAA S1 standard. Each SEP event in the CDAW-SEP list, the associated flares and CMEs, and their properties are identified when available (Gopalswamy et al. 2003a; 2015). All the information is compiled and extended from an earlier report by Gopalswamy (2003; 2012). SEP events from SC 23 & 24 are studied by Gopalswamy et al.(2004a; 2004b; 2014), Mäkelä et al. (2015), Thankur et al. (2016) and Xie et al. (2016).

## 4. GSEP EVENTS LIST

The preliminary data processing and work structure in the integration and development of the catalog is illustrated in Figure 1.

### 4.1. *GOES Data*

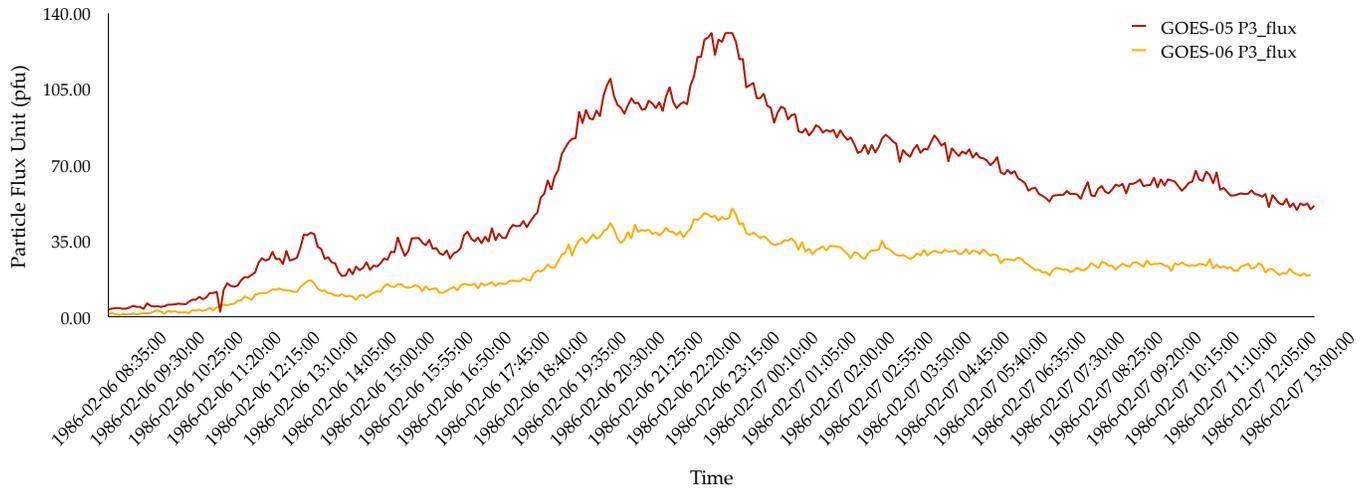

**Figure 2.** Time series of an SEP event showing the variation in the fluxes captured by the GOES-05 ("secondary") and GOES-06 ("primary") satellites.

We utilize the integral proton fluxes measured on board GOES 05–15 that are archived on the NOAA website[4]. Lower energy fluxes corresponding to P1 channel were not used because of their high sensitivity to interplanetary disturbances. We performed a visual inspection of GOES data to understand flux enhancements and identify the more accurate observational sources. Although the design of EPS and EPEAD onboard GOES has not changed, some variations in the measurements have occurred between satellites. As the instruments were built with passive shielding, measurements are affected by significant side and rear-penetration effects, i.e., particles can pass through the shielding from any direction and be counted as though they had entered through the nominal detector entrance aperture (Posner 2007; Bruno 2017). This is crucial as the differences in sensor data could impact the identification of an event and its timing in cases where the flux levels are near the event threshold. For instance, in the intercalibration of GOES 8–15 solar proton detectors by Rodriguez et al. (2014), it was reported that the relative responses between GOES primary and secondary agree to within ±20% while varying during a significant event. Based on a calibration study of the EPSs onboard GOES-5, -7, -8, and -11, Rodriguez et al. (2017) validated the derived cross-calibrated energies by comparison with the STEREO data. In their study, they utilize the integrated proton fluxes calculated using the algorithm developed by R. Zwickl (unpublished, 1989). See Rodriguez et al. (2017); Appendix, for the details of the algorithm.

In light of this significant/known instrument-to-instrument variation, we carefully identified reliable missions by comparing the time intensity plots of primary and secondary GOES instruments for all the observed event periods. To illustrate, Figure 2 shows the difference in the flux level enhancements between primary (GOES-06) and secondary (GOES-05) satellites for an SEP event during SC 22. Therefore, it is not a straightforward option to utilize data from the primary GOES satellite. The differences in the measurements of solar proton fluxes between GOES primary and secondary are due to the geomagnetic cut-off, i.e., the effect of variation of the magnetic field configuration with geomagnetic longitude (Rodriguez et al. 2014). Therefore, we consider the strongest proton signal as the best for two reasons; (1) the corrected fluxes have been checked with intercalibration and (2) the peak values of the strongest signals closely match with those reported in the CDAW-SEP and NOAA-SEP lists. We performed additional data processing which ensured a compromise with imputable data gaps on better sensory responses of the instruments. That is, if the primary (secondary) GOES has a better response, but with more data gaps than the secondary (primary) GOES, then we consider the GOES primary (secondary) as a reliable data provider. We impute all the data gaps with linear interpolation. In addition, the EPEAD data were inspected for differences in enhancements between the "East" and "West" channels. According to Rodriguez (2010), the East-West effects are more relevant at lower energies. We have examined all SEP event temporal profiles and observed up to ±190% (in SC 22), ±90% (in SC23) and ±30% (in SC 24) differences between primary and secondary source energy channels.

---

[4] https://www.ngdc.noaa.gov/stp/satellite/goes/index.html

**Table 2.** Header description in the GSEP list

| Header | Description |
| --- | --- |
| sep_index | Index for the GSEP events list |
| pp_id | Event identifier in the PSEP catalog |
| cdaw_sep_id | Event identifier in the CDAW-SEP list |
| timestamp | Start time of the event in PSEP |
| cdaw_start_time | Start time of the event in CDAW-SEP |
| cdaw_max_time | Event peak time in CDAW-SEP |
| cdaw_evn_max | Event peak flux in CDAW-SEP |
| cme_id | Identifier of the CME in LASCO CME catalog |
| cme_launch_time | Start time of the CME |
| cme_1st_app_time | First appearance time of the CME |
| lasco_cme_width | Width of the CME in Lasco Catalog |
| p_cme_width | Width of the CME in PSEP |
| lasco_linear_speed | CME velocity reported by LASCO |
| p_cme_speed | CME velocity reported by PSEP |
| fl_id | Autogenerated unique flare identifier |
| fl_start_time | Start time of the flare |
| fl_peak_time | Time of Flare Maximum |
| fl_rise_time | Time taken to reach peak |
| fl_lon | Longitude of flaring region |
| fl_lat | Latitude of flaring region |
| fl_goes_class | GOES Flare classification |
| noaa_ar | NOAA Active region number |
| noaa_ar_uncertain | Flag for NOAA active region number uncertainty |
| harpnum | HMI Active Region Patch (HARP) number corresponding to the originating NOAA AR |
| noaa_pf10MeV | Peak flux in the NOAA-SEP >10 MeV channel |
| ppf_gt10MeV | Peak flux in the PSEP >10 MeV channel |
| ppf_gt30MeV | Peak flux in the PSEP >30 MeV channel |
| ppf_gt60MeV | Peak flux in the PSEP >60 MeV channel |
| ppf_gt100MeV | Peak flux in the PSEP >100 MeV channel |
| fluence_gt10MeV | Peak fluence in the PSEP >10 MeV channel |
| fluence_gt30MeV | Peak fluence in the PSEP >30 MeV channel |
| fluence_gt60MeV | Peak fluence in the PSEP >60 MeV channel |
| fluence_gt100MeV | Peak fluence in the PSEP >100 MeV channel |
| gsep_pf_gt10MeV | Peak flux in the GSEP >10 MeV channel |
| gsep_max_time | Event peak time in GSEP |
| m_type2_onset_time | metric type II radio burst start time |
| dh_type2_onset_time | decameter-hectometric (DH) type II radio burst start time |
| noaa-sep_flag | 1 if present in NOAA-SEP; 0 otherwise |
| Inst_category | GOES Instrument category: P for Primary; S for Secondary |
| Comments | Retained from PSEP |
| Notes | Retained from PSEP |
| Fe_e_p_shock_notes | Retained from PSEP |
| gsep_notes | Data observational notes, if any |
| slice_start | Start time of the slice |
| slice_end | End time of the slice |

NOTE—The catalog available at Harvard Dataverse: 10.7910/DVN/DZYLHK

### 4.2. *Integration of catalogs*

The SEP events from the PSEP and the CDAW-SEP catalogs, are integrated into Geostationary Solar Energetic Particle (GSEP) events list of this paper. In addition, we utilize the NOAA-SEP list as a reference catalog. That is, each event in the GSEP list is cross-checked with the reference catalog. Furthermore, a binary secondary source verification indicator is given in the metadata, where 0 represents no source was found in the NOAA-SEP list and 1 if found.

The PSEP and CDAW-SEP catalogs contain many valuable parameters related to temporal characteristics, integrated flux information and solar source metadata. Details on the SEP events' start time, peak time, and peak flux value in the >10 MeV channel are provided. They also report associations of SEP events with a parent solar eruption. Information such as the event coordinates about the associated flare and CME is provided as well. This information is used to determine if the entries in the PSEP, CDAW-SEP and NOAA-SEP catalogs represent the same event, notwithstanding the minor differences in temporal characteristics, i.e., if they happened simultaneously with the same enhancements or if they vary and are eventually different events.

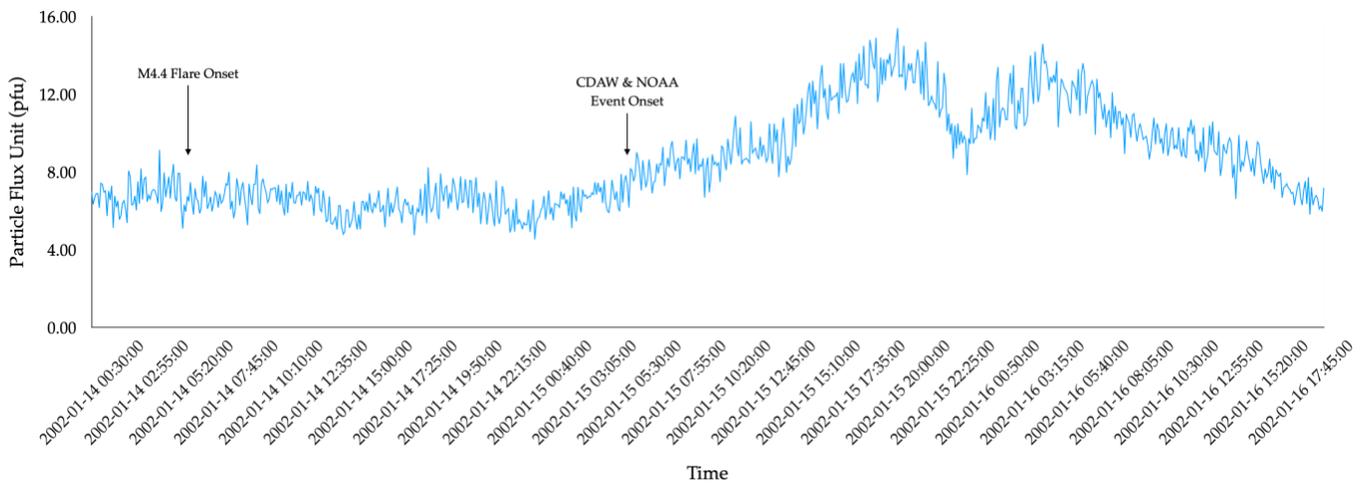

**Figure 3.** Time profile of an SEP event occurring between 14 and 16 January 2002. The time series begins with the event start time (2002-01-14 00:30:00) according to the PSEP catalog. A solar flare of magnitude M4.4 occurs at 05:30:00 followed by a CME six minutes later from the western limb. The second arrow points to the onset of the SEP event (2002-01-15 05:35:00) as considered by the CDAW-SEP and NOAA-SEP catalogs. The event start time differs by >29 hours compared to PSEP. For this event, both catalogs report the same solar source (flare and CME). We retain the event onset as reported in the CDAW-SEP list.

### 4.3. *Challenges*

Key challenges we had to address in integrating multiple catalogs were: overlapping events, repetitions, and different criteria in event start time, peak time and the corresponding peak fluxes. Different catalogs implemented different data calibration methods as well. To illustrate with an example, the different onset criteria, a time series plot of GSEP event 211 is shown in Figure 3. Here, PSEP considers the event onset prior to a M4.4 class solar flare and >29 hours ahead compared to the CDAW-SEP (2002-01-14 00:30:00 and 2002-01-15 05:35:00, respectively). The associated flare had a rise time of 58 minutes and is followed by a CME erupting behind the western limb. We take into consideration the start time as reported in CDAW-SEP as it accounts for the SWPC threshold of a significant SEP event. The event peak is observed on 2002-01-15 at 20:00:00 to reach a maximum of 15 pfu in the >10MeV channel.

In the above example, although both source catalogs refer to the same event, the difference in start time is due to the criterion (like, the event threshold) in considering a distinct onset. Such dissimilarities between catalogs have been verified with the time profiles. Also, plots available by NOAA[5] were used to cross-verify our time series plots and conclude whether an event occurred or crossed the NOAA threshold on a specific date.

---

[5] https://satdat.ngdc.noaa.gov/sem/goes/data/plots/

### 4.4. Description of the Catalog

Our integrated catalog gathers SEP event records from multiple sources and provides relevant metadata useful for space weather research. The headers in the GSEP list and their descriptions are presented in Table 2. The majority of the SEP events have been captured by the primary GOES instrument. However, a total of nineteen events were observed by the secondary instrument. Among them thirteen events are in SC 22, and three for each of SCs 23 and 24. The final catalog has observed source instrument flag "P" or "S" indicating whether the event was measured in the primary or secondary GOES instrument, respectively.

### 4.5. Time series slices

The plots of time series slices from the GSEP list consist of:

- Electron fluxes (channels: E2 and E3 i.e., >2.0 MeV and >4.0 MeV).
- Proton fluxes (channels P2 to P7 i.e., from >5.0 MeV to >100.0 MeV).

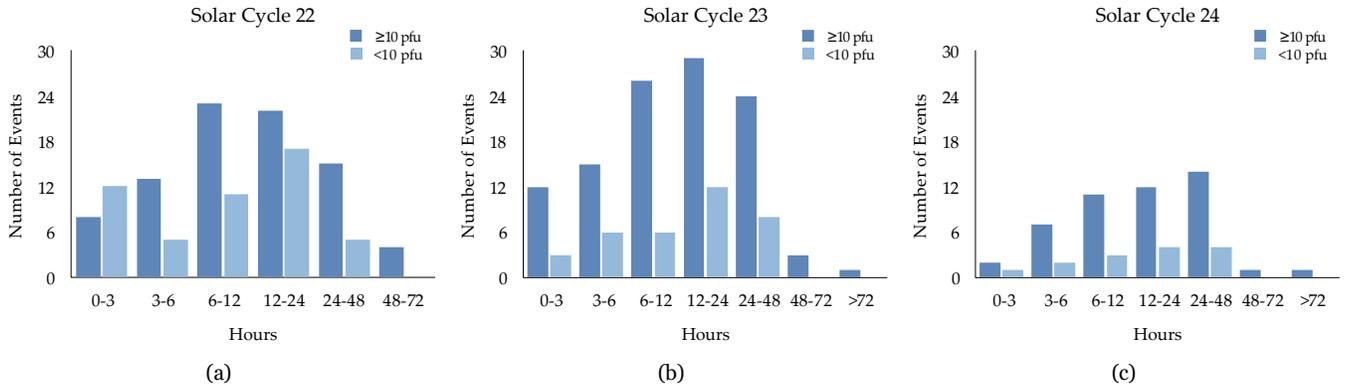

**Figure 4.** Histograms of rise times (i.e., times elapsed between onset and peak flux) for the GSEP events. Shown are numbers of SEP events vs. their rise times, classified in bins of 0 - 3, 3 - 6, 6 - 12, 12 - 24, 24 - 48, 48 - 72, and >72 hours. Weak enhancements (<10 pfu for protons >10 MeV) and events above the NOAA S1 scale are included in (a-c) for SCs, 22, 23 and 24, respectively.

Here, the integral fluxes are derived from the observations of GOES/EPS from 1986 to 2012 and GOES/EPEAD from 2013 to 2017. Each value in a time series data represents a 5-minute interval. The length of each time profile denotes the events' start and end times. These characteristics, which help describe the flux evolution and the data quality, provide visual information for selecting events for further analysis.

To summarize, we downloaded the GOES integral fluxes and classified the data into respective instruments. We visually inspected the primary and secondary observations to understand the overlaps, data gaps and intensity variations. We merged the data in series for each SC. Finally, we sliced the GOES particle fluxes with reference to the onset/start time and the observed end time of each SEP event as defined in the GSEP list. The identification metadata is encoded in the filenames of time series data instances. We use the SEP event initiation as a default reference time. The best and simplest form we opted for contains the event date and time that correspond to the timestamp of the event onset; for example, 2017-09-10 04-25.csv). We also assign and maintain the correspondence between our index and the indices in the primary source catalogs.

## 5. RESULTS

Integrating the primary catalogs, we have obtained 335 unique events. However, after comparing with the reference one, six more events were included. In the course of this work, we discussed with NASA's Space Radiation Analysis Group (SRAG) to validate our SEP events list. Hence, a total of 341 events are available in the GSEP catalog from 1986 to 2017. The time series profile for each event has been visually inspected for confirming the event definition. Of these, 96 events fall under the weak enhancement category (peak flux <10 pfu at >10 MeV), while 245 events achieve a peak flux >10 pfu in the >10 MeV channel over the past three SCs. In Table 3, the number of events is provided according to the flux enhancements in different levels of the NOAA solar radiation storm scale. Here, S0 is a custom scale used to denote a sub-event of proton fluxes below 10 pfu. As the PSEP catalog has no records beyond 2013, the number of sub-events in the GSEP catalog for SC24 is lesser as compared to the prior two cycles.

An important property we want to address is the timescales of SEP events. We retained the criteria followed by the source catalogs; PSEP and CDAW-SEP. During SC22, all events' start times are based on PSEP. For cycles 23 and 24, we choose CDAW-SEPs' criteria and switch to PSEP or NOAA-SEP when discrepancies occur. That is, for an event, whenever there is a CDAW-SEP onset available, it is visually verified to see if the temporal profile matches the event definition. If yes, then the CDAW-SEP onset is used. Else, we prefer PSEP. In Figure 4(a-c), we present rise time distributions for both weak enhancements and strong events (S1 and above) for the three latest solar cycles. It can be seen that the majority of the events last more than six hours over the rising phase, and several events take more than 24 hours to reach peak proton fluxes. A great number events have the rising phase predominant anywhere from 24 to 48 hours while some events take more than 48 hours to reach peak proton fluxes. Interestingly, this trend has reduced with the solar cycle. In addition, a few large events in the last two solar cycles take more than 72 hours to reach peak fluxes.

In some cases with complicated SEP event temporal profiles, the peak flux occurs after an initial, or a pair of, peaks. Although initial peaks could directly reflect the parent (SF or CME) properties, later peaks may be due to particle transport effects (Kihara et al. 2020). Regardless, it is the proton fluence (i.e., the time-integrated flux) that determines doses which are crucial when space weather effects are considered.

In the GSEP list, the NOAA active region number is available for 297 SEP events. Full source information (i.e., a flare and a CME) exists for 164 events. Nonetheless, 145 SEP events are associated with flares-only, and 24 SEP events are associated with CME-only. Because the necessary CME data is unavailable for SC 22, the numbers for CME-SEP association are less. For 309 SEP events where a source flare could be determined, 84 events are weak and 225 events are large. There are two SEP events with no recorded flare peak time; five SEP events with no recorded GOES flare class; and twelve SEP events (nine in SC 22 and three in SC 23) with no reference to the flare location. There are eight SEP events (three in SC 22 and five in SC 23) in the GSEP list that do not have any source association. Three of these events have weak proton enhancements. Among the five large SEP events, three are flagged as probable Energetic Storm Particle (ESP) events by SRAG.

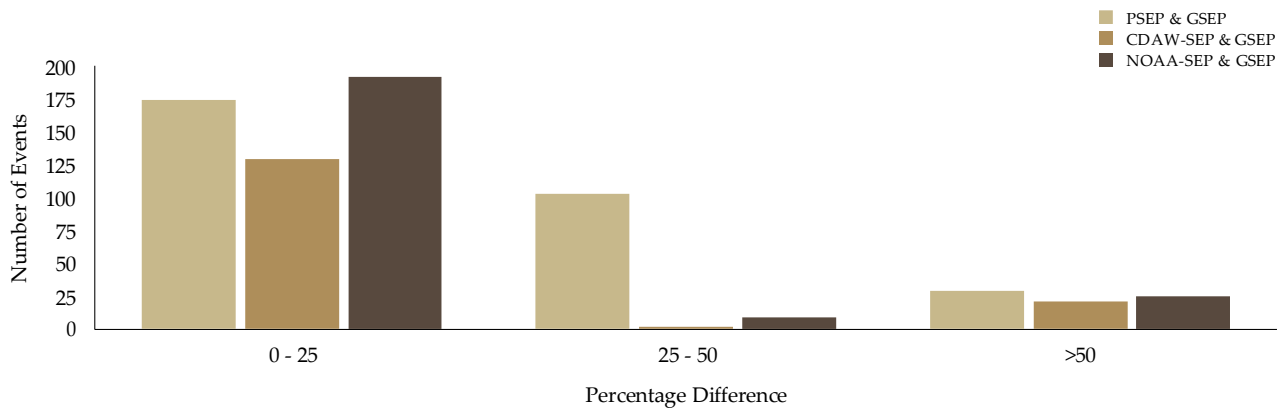

**Figure 5.** Distribution of the absolute percentage differences classified in bins of 0 - 25, 25 - 50, and >50 percent in the peak proton fluxes between GSEP and (a) PSEP, (b) CDAW-SEP & (c) NOAA-SEP.

In terms of sensory response to measure protons, the peak flux recorded by the primary GOES instrument is better than secondary in the majority of the cases. Nonetheless, the peak values reported by PSEP, CDAW-SEP and NOAA-SEP differ from GSEP metadata on several occasions. In Figure 5, the distribution of percentage difference comparing the GSEP list with the PSEP and CDAW-SEP catalogs is shown.

A comparative summary between the GSEP and the PSEP and CDAW-SEP catalogs is given below:

- GSEP and PSEP lists:

1. 280 out of 303 events from the PSEP catalog are within ±50% difference in the peak flux enhancements with respect to GSEP.

2. There are nine events where PSEP records a peak flux of <10 pfu, but GSEP records >10 pfu. Out of these, five events are close enough with fluxes between 9 to 12 pfu, while differences in the remaining four events are significant. All these events are listed in Table 4, Section A of the Appendix.

3. PSEP event 185 (psep185) reports same proton fluxes as event 186 (psep186). This could be a possible computational or human error because the episode appears to be entangled. It is a relatively weak event and of short duration. However, we did not merge the two events because they are associated with distinct solar sources, both flares and CMEs. (See Section B of the Appendix.)

- GSEP and CDAW-SEP lists:

  1. 126 out of 150 events have peak flux enhancements agreeing within ±20%.
  2. Slight discrepancies exist for extremely large events (with peak proton fluxes at least >1500 pfu).
  3. Due to the variation in the event identification criteria, some of the event peaks have been missed by the CDAW-SEP.

- GSEP and NOAA-SEP lists:

  1. 189 events are within ±20% difference in the peak flux enhancements.
  2. Seven events with higher differences correspond to extremely large events (of peak proton fluxes at least >1500 pfu).

**Table 3.** Number of SEP events with respect to the NOAA solar radiation storm scale in the 10 MeV channel across the last three solar cycles.

| Scale (flux level in pfu) | SC22 | SC23 | SC24 |
|---|---|---|---|
| S0*(<10) | 48 | 34 | 14 |
| S1 (≥10 to <$10^2$) | 49 | 61 | 31 |
| S2 (≥$10^2$ to <$10^3$) | 21 | 31 | 10 |
| S3 (≥$10^3$ to <$10^4$) | 14 | 13 | 6 |
| S4 (>$10^4$) | 3 | 6 | 0 |
| Total | 135 | 145 | 61 |

NOTE—*S0 is a custom label to indicate a sub-event.

## 6. CONCLUSIONS

We present an integrated Geostationary Solar Energetic Particle Events Catalog (GSEP) created from a set of available SEP event catalogs based on the particle fluxes of GOES missions from 1986 to 2017. We homogenized the SEP events from two primary catalogs (Papaioannou et al. (2016) and CDAW-SEP) by filtering all events, i.e., removing overlapping and repetitive episodes. Then we cross-checked the SEP events with the reference, i.e., the NOAA-SEP list. Every entry in the catalog is assigned a new index for SEPs with reference to the indices of the source catalogs. The metadata provides an association of an SEP event to the corresponding source solar eruption, where available. The main summary of the paper is as follows:

1. There are 341 SEP events in the GSEP list. In that, 245 events have peak proton fluxes >10 pfu in the >10 MeV channel.

2. The particle fluxes of each event are visually inspected for errors and variations by parallel comparison of time profiles .

3. The fluxes are further sliced with respect to the event start and end times as reported in the GSEP metadata.

4. The headers in the GSEP list describe physical descriptors (both those stored in the source catalogs and calculated by us) and carry relevant indicators (data quality, observed GOES instrument, and parallel reports.)

5. The time series slices are published as a data set to implement Machine Learning or other statistical analysis for experimenting on SEP event forecasting.

This work provides a catalog from which users can explore SEP events with parameters of interest for various statistical studies and Machine Learning exercises. Also, it provides a reference to various parameters for each event, allowing researchers to understand if the event satisfies the criteria for case studies. Our approach is to contribute to the SEP research community with a combined database and present additional data for each event. The integrated GSEP catalog provides a one-stop database for researchers to study SEP events using an extensive, long-term data archive.

Our GSEP dataset is available at Harvard Dataverse: 10.7910/DVN/DZYLHK. The plots and statistics presented in this study is based on version 4.0 of the dataset.

We acknowledge the use of data from NOAA-GOES missions and thank the team for the availability of particle data. We also thank the teams behind the catalogs: PSEP, CDAW-SEP, and NOAA-SEP; for the opportunity to utilize their work. Author Petrus Martens' contribution is supported by NASA SWR2O2R Grant 80NSSC22K0272 . Author S. Rotti carried out this work while supported by the NASA FINESST Grant 80NSSC21K1388. SR thanks (1) Dr. Hazel Bain of NOAA for information on the GOES primary and secondary observations. (2) Dr. Steve Johnson of NASA-SRAG for discussing much of his work in detail and agreeing to merge the efforts. The explanations on the events of different characters were crucial to classify and flag the SEP events. We thank the anonymous reviewer for constructive comments on the manuscript that improved the contents of the paper.

## APPENDIX

### A. WEAK EVENTS IN PSEP

In Table 4 below, the nine events that are reported in PSEP with peak proton fluxes below 10 MeV are listed. The index refers to the event number in the GSEP list. The next column indicates the event onset followed by event maximum timestamp as reported in NOAA-SEP and GSEP. The last three columns show the peak proton fluxes (in pfu) from the PSEP, NOAA-SEP and GSEP lists, respectively.

**Table 4.** SEP events reported in PSEP with peak fluxes <10 pfu but observed to be >10 pfu in the GSEP list.

| sep_index | event_start_time | noaa_max_time | gsep_max_time | ppf_gt10MeV (PSEP) | noaa_pf10MeV (NOAA-SEP) | gsep_pf_gt10MeV (GSEP) |
|---|---|---|---|---|---|---|
| **gsep_034** | 1989-06-18 15:00:00 | 1989-06-18 19:10:00 | 1989-06-18 20:25:00 | 9.24 | 18 | 10.8 |
| **gsep_058** | 1989-11-15 07:05:00 | 1989-11-15 09:10:00 | 1989-11-15 09:05:00 | 4.85 | 71 | 38.3 |
| **gsep_062** | 1990-03-28 13:50:00 | 1990-03-29 10:05:00 | 1990-03-29 10:05:00 | 2.14 | 16 | 15.9 |
| **gsep_086** | 1991-03-31 21:25:00 | – | 1991-04-03 09:10:00 | 3.04 | – | 25.5 |
| **gsep_117** | 1992-03-16 04:35:00 | 1992-03-16 08:40:00 | 1992-03-16 09:00:00 | 9.11 | 10 | 10.4 |
| **gsep_130** | 1993-03-06 21:15:00 | – | 1993-03-07 07:10:00 | 9.73 | – | 10.8 |
| **gsep_195** | 2001-09-15 12:20:00 | 2001-09-15 14:55:00 | 2001-09-15 14:55:00 | 9.49 | 12 | 11.6 |
| **gsep_200** | 2001-10-19 17:45:00 | 2001-10-19 22:30:00 | 2001-10-19 22:30:00 | 9.53 | 12 | 11.7 |
| **gsep_295** | 2011-10-22 12:15:00 | – | 2011-10-23 15:35:00 | 7.88 | – | 13.1 |

## B. ENTANGLED SEP EVENTS

In Figure 6, the time profiles of two SEP events (182 and 183) in the GSEP list are shown. The latter is a very large event, while the former is apparently a weak event with peak proton flux of 4 pfu at >10 MeV. On 2001-04-02, the first SEP event appears associated to an X1.1 flare at 10:58:00, while an X20 flare at 21:32:00 leads to the second SEP event. The flaring active region (9393) is positioned at the western hemisphere of the Sun while erupting. Both the SEP events are associated with distinct CMEs, detected after the respective flares.

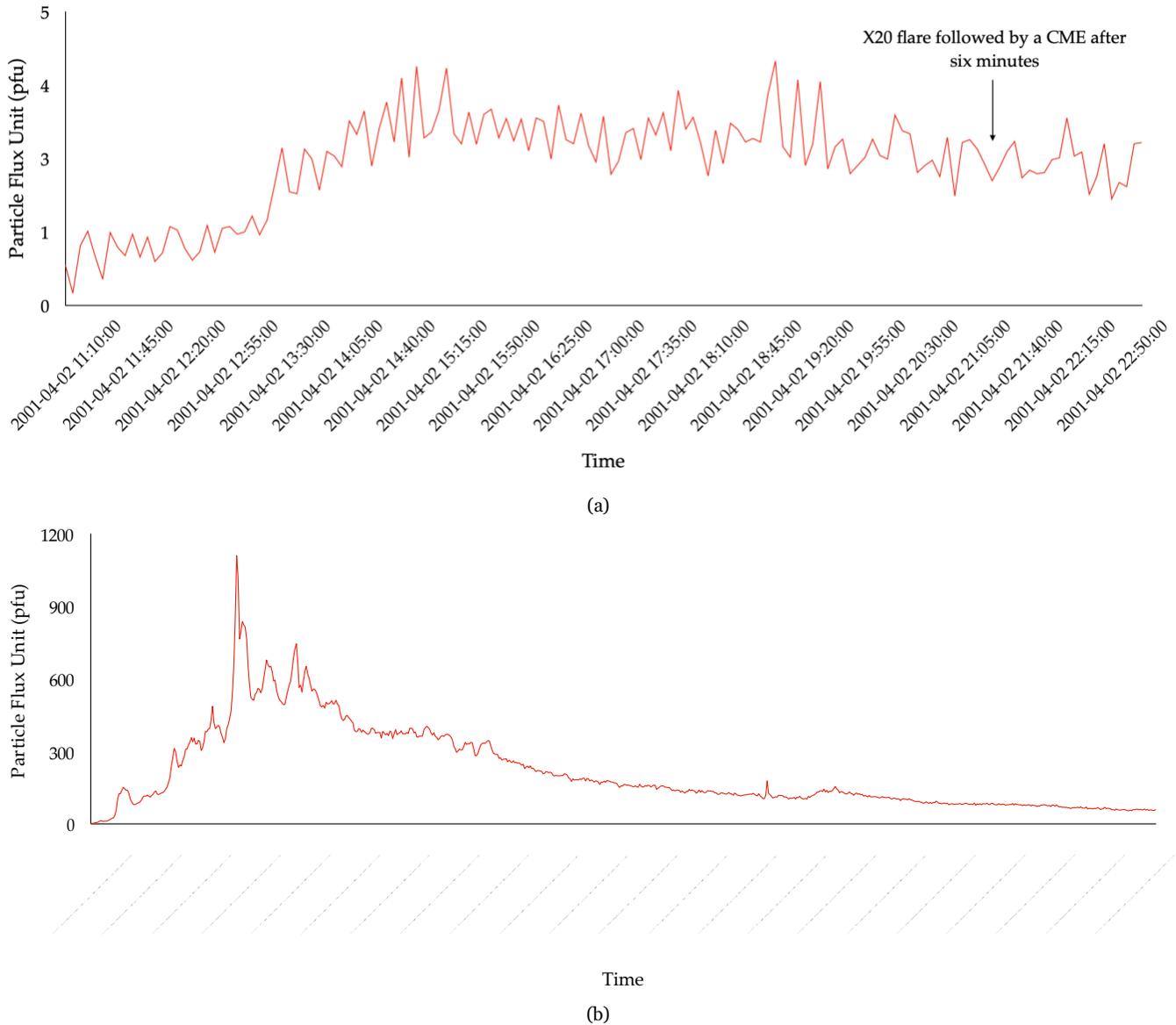

**Figure 6.** Time series profiles of GSEP event 182 in (a) and event 183 in (b). The first SEP event was due to an X1.1 flare and a CME (2001-04-02 10:58:00), while the second event was triggered due to an X20 flare (2001-04-02 21:32:00) and a CME erupted after six minutes. Both events originated from NOAA AR 9393.